\documentclass[review]{elsarticle}

\usepackage{amsmath}
\usepackage{amssymb}
\usepackage{hyperref}
\usepackage{graphics}      
\usepackage{graphicx} 
\usepackage{bm}
\usepackage[english]{babel}
\usepackage[utf8]{inputenc}

\journal{Journal of \LaTeX\ Templates}

\begin{document}

\begin{frontmatter}

\title{Magnetic and thermodynamic properties of the octanuclear nickel phosphonate-based cage}

\author{Hamid Arian Zad\corref{mycorrespondingauthor}}
\address{A.I. Alikhanyan National Science Laboratory, Alikhanian Br. 2, 0036 Yerevan, Armenia}
\address{ ICTP, Strada Costiera 11, I-34151 Trieste, Italy}
\ead{arianzad.hamid@yerphi.am}

\author{Ralph Kenna}
\ead{csx267@coventry.ac.uk}
\address{Applied Mathematics Research Centre, Coventry University, Coventry, CV1 5FB, England, UK}

\author{Nerses Ananikian}
\address{A.I. Alikhanyan National Science Laboratory, Alikhanian Br. 2, 0036 Yerevan, Armenia}
\address{CANDLE Synchrotron Research Institute, Acharyan 31, 0040 Yerevan, Armenia}
\ead{ananik@yerphi.am}


\begin{abstract}
We report a detailed theoretical investigation into the influence of anisotropy on the magnetic and thermodynamic properties of an octanuclear nickel phosphonate cage with  butterfly-shaped molecular  geometry, namely\\
$\mathrm{Ni}_8(\mu_3-\mathrm{OH})_4(\mathrm{OMe})_2(\mathrm{O}_3\mathrm{PR}_1)_2 (\mathrm{O}_2\mathrm{C}^t\mathrm{Bu})_6  (\mathrm{HO}_2\mathrm{C}^t\mathrm{Bu})_8$. 
To validate our exact diagonalization approach, we firstly  compare  results with simulations and experiment in the isotropic case. 
Having established concurrence, we then  introduce uniaxial single-ion anisotropy and Heisenberg exchange anisotropy between interacted nickel atoms. 
We then examine effects of  both anisotropy parameters on the magnetization process, as well as on the specific heat of the model.
We predict intermediate magnetization plateaus, including zero plateau, and magnetization jumps with magnetic ground-state phase transitions at low temperature $T=1$K. 
The magnetization plateaus are strongly dependent on both the levels of exchange anisotropy and  single-ion anisotropy. 
Varying the former leads to change in width and magnetic position of all intermediate plateaus while they become wider upon increasing the latter.
The specific heat of the model manifests a Schottky-type maximum at moderate temperature in the presence of weak magnetic fields, when the system is isotropic. 
The introducion of aniostropy results in  substantial variations in the thermal behavior of the specific heat. 
Indeed, by tuning anisotropy parameters the Schottky peak convert to a double-peak temperature dependence that coincided with the magnetization jumps.
We call for these theoretical predictions to be verified experimentally at low temperature.
\end{abstract}


\end{frontmatter}


\section{Introduction}
Low-dimensional quantum systems with competing interactions and geometrical frustration are of continued interest in condensed matter physics, material science and chemistry \cite{Kahn,Carlin,Furrer,Lacroix,Sachdev}. 
The interest in the magnetic and thermal properties of  metal-containing compounds of this type stem not least from their manifestation of various ground-state phase diagrams at zero and low temperatures   \cite{Sachdev,Kitaev,Gu,Dillenschneider,Ivanov,Werlang,Rojas2,RojasM,Strecka16}. 
For instance, experimental investigations into the nickel homometallic magnetic compound $[\mathrm{Ni}_3(\mathrm{fum})_2(\mu_3-\mathrm{OH})_2(\mathrm{H}_2\mathrm{O})_4]_n\cdot (2\mathrm{H}_2\mathrm{O})_n $ indicate the coexistence of both antiferromagnetic and ferromagnetic interactions \cite{Konar2002}. 
These results inspire analytical studies of magnetic properties on a diamond chain for low temperatures \cite{Ananikian2014,Ananikian2016,Abgaryan2}.

Single-molecule magnets have also attracted increased interest owing to the fact that  they can possibly be characterized by Heisenberg models 
 \cite{Sheikh,Sheikh2016,Breeze,Khuntia1111,Oyarzabal15,Kalita2018,str17a,kar17,str18b,strPhysB2018,strPhysRevB2018,Arian20191}. 
Polynuclear complexes involving transition-metal and lanthanide-metal ions, synthesized from either identical or different metal ions, are amongst the themes of increasing interest in the study of molecular magnetism. 
In the past two decades, it has become possible to synthesize a large variety of compounds including identical transition-metal ions with the geometry of butterfly-shaped subunit structures. 
These can be properly introduced in terms of Heisenberg models such as:  
$\big[\mathrm{Fe}^\mathrm{III}_6(\mu_3-\mathrm{O})_2(\mu-\mathrm{OH})_2\{\mu-(\mathrm{C}_6\mathrm{H}_{11})_2\mathrm{PO}_2\}_6(\mu-t\mathrm{BuCO}_2)_6(\eta_1-\mathrm{OH}_2)_2]\cdot 2\mathrm{CH}_3\mathrm{CN}\cdot \mathrm{CH}_2\mathrm{Cl}_2\big]$ (for more details see Ref. \cite{Oyarzabal15}), and nickel containing complex
 $\big[\mathrm{Ni}_8(\mu_3-\mathrm{OH})_4(\mathrm{OMe})_2(\mathrm{O}_3\mathrm{PR}_1)_2 (\mathrm{O}_2\mathrm{C}^t\mathrm{Bu})_6 (\mathrm{HO}_2\mathrm{C}^t\mathrm{Bu})_8\big]$ \cite{Sheikh,Sheikh2016,Breeze}. Also, several nickel containing compounds consisting of mixed transition-metal ions have widely been synthesized; examples include heterometallic octanuclear 
$\mathrm{Ni}^{\mathrm{II}}_4\mathrm{Ln}^{\mathrm{III}}_4$ (Ln = Y, Gd, Tb, Dy, Ho, and Er) complexes containing two butterfy-shaped $\mathrm{Ni}^{\mathrm{II}}_2\mathrm{Ln}^{\mathrm{III}}_2\mathrm{O}_4$ \cite{Kalita2018},
as well as heterometallic $(3 \times 3)-$grid $\mathrm{M}^{\mathrm{II}}\mathrm{Cu}^{\mathrm{II}}_4\mathrm{Cu}^{\mathrm{I}}_4$ (M = Ni, Cu, and Zn) reported in Ref. \cite{Bao2013}.

Quantum fluctuations play the essential role for phase transitions at zero temperature and may therefore describe the nature of materials in the low-temperature real world.
Therefore quantum phase transitions in various spin models \cite{Arian20192,Arian1,Feng2007,Saadatmand,Valverde,Ananikian2012,Abgaryan1,Strecka1,Arian2} have been amongst the most interesting topics to study in statistical mechanics. 
Further studies of quantum spin models have provided precise predictions for ground-state phase transitions in the presence of external magnetic fields,  which can be induced through  exchange couplings  \cite{Sahoo1,Sahoo2,Giri,Hovhannisyan}. 
Research attention currently focuses on zero- and low-temperature magnetization curves of small spin clusters consisting of transition-metals and intriguing features such as fractional magnetization (quasi)plateaus, magnetization jumps and magnetization ramps \cite{str17a,kar17,str18b,strPhysB2018,strPhysRevB2018} add to the interest.  

The aim of this work is to investigate how anisotropy effects can substantially  modify or control the ground-state properties of magnetic systems \cite{Furrer2011,Xiang2011}. 
To this end we examine the low-temperature magnetization process and the specific heat of the octanuclear nickel phosphonate cage \\
$\big[\mathrm{Ni}_8(\mu_3-\mathrm{OH})_4(\mathrm{OMe})_2(\mathrm{O}_3\mathrm{PR}_1)_2 (\mathrm{O}_2\mathrm{C^tBu})_6(\mathrm{HO}_2\mathrm{C^tBu})_8\big]$ in the presence of an external magnetic field with additional anisotropic terms.
These are the Heisenberg exchange anisotropies $\Delta_1$ and $\Delta_2$ and single-ion anisotropy $D$. 
To investigate the low-temperature magnetization behavior and isothermal specific heat of the nickel complex discussed above, we utilize exact numerical methods to diagonalize the Hamiltonian of the model. 
To test for correctness and completeness, we compare our exact results of the magnetization process with those obtained through Quantum Monte Carlo (QMC) methods. 
We use the subroutine ${dirloop-sse-}$, a package from the Algorithms and Libraries for Physics Simulations (ALPS) project. 
This provides a full generic implementation of the QMC procedure called the directed loop algorithm in the Stochastic Series Expansion representation \cite{Bauer2011,Albuquerque2007}. 

\begin{figure}
\begin{center}
\resizebox{0.5\textwidth}{!}{%
  \includegraphics{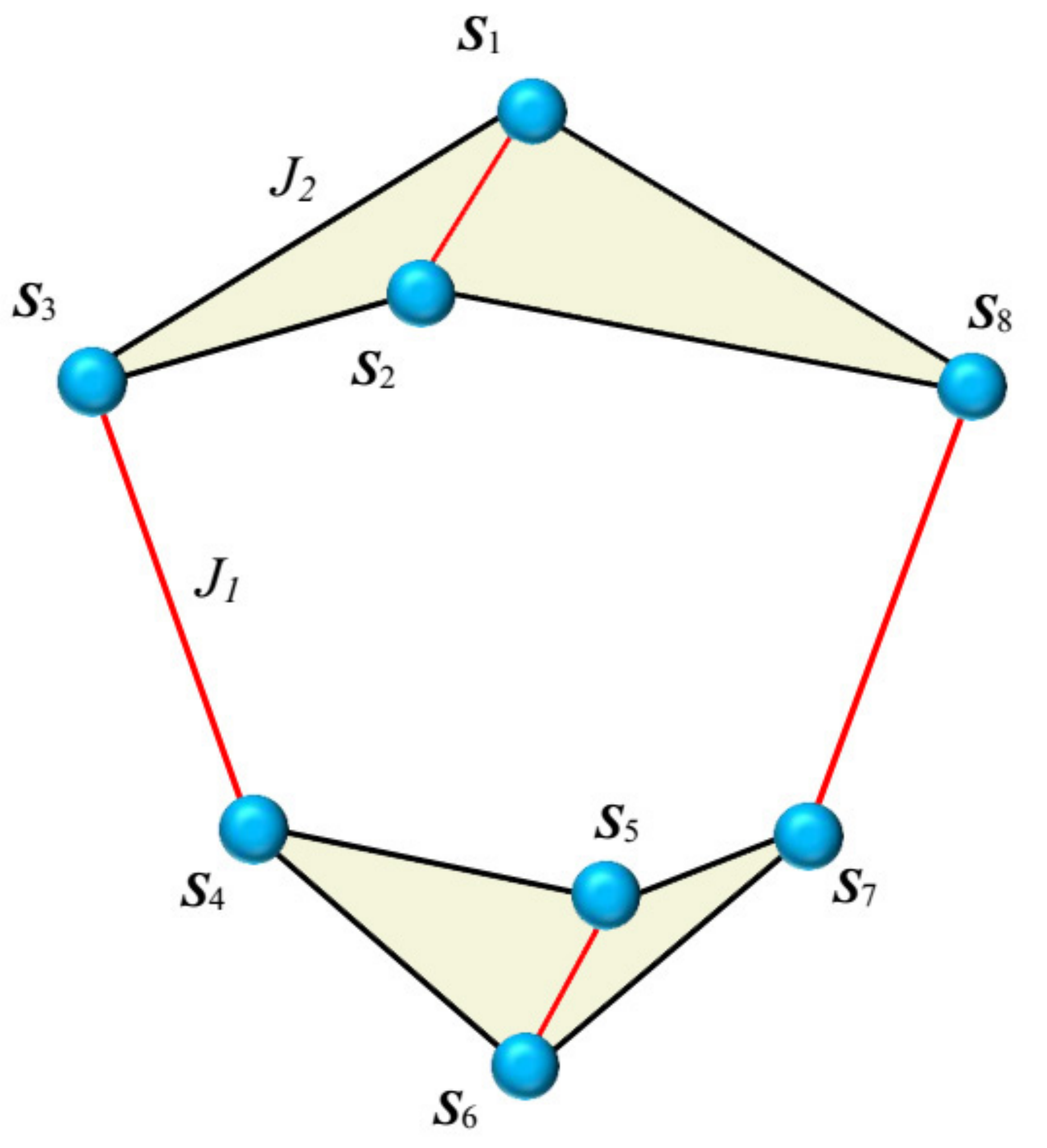}
  }
\caption{Schematic structure of the Heisenberg octanuclear nickel phosphonate cage. The balls denote Ni atoms that are connected together via Wing-Body (WB) interadimer interaction $J_2$ and the corresponding Body-Body (BB) interdimer interaction $J_1$.}
\label{fig:Model}
\end{center}
\end{figure}

The paper is organized as follows. 
In Sec. \ref{Model} we introduce the exactly solvable model. 
In Sec. \ref{TP}, we discuss the low-temperature magnetization process and specific heat of the model under consideration by assuming anisotropy properties. 
Conclusions and future outlooks are discussed in Sec. \ref{conclusions}.

\section{Model}\label{Model}
 Hamiltonian of the octanuclear nickel phosphonate-based cage  $\big[\mathrm{Ni}_8(\mu_3-\mathrm{OH})_4(\mathrm{OMe})_2(\mathrm{O}_3\mathrm{PR}_1)_2 (\mathrm{O}_2\mathrm{C}^t\mathrm{Bu})_6 (\mathrm{HO}_2\mathrm{C}^t\mathrm{Bu})_8\big]$ shown in Fig. \ref{fig:Model} can be expressed as
\begin{equation}\label{Hamiltonian}
\begin{array}{lcl}
H=-J_1 \big[\boldsymbol{S}_{1}\cdot\boldsymbol{S}_{2} + \boldsymbol{S}_{3}\cdot\boldsymbol{S}_{4}+\boldsymbol{S}_{5}\cdot\boldsymbol{S}_{6}+\boldsymbol{S}_{7}\cdot\boldsymbol{S}_{8}\big]\\
\quad\quad-J_2 \big[\boldsymbol{S}_{1}\cdot\boldsymbol{S}_{3} + \boldsymbol{S}_{2}\cdot\boldsymbol{S}_{3}+\boldsymbol{S}_{4}\cdot
\boldsymbol{S}_{5}+\boldsymbol{S}_{4}\cdot\boldsymbol{S}_{6}+\\
\quad\quad\quad\quad \boldsymbol{S}_{5}\cdot\boldsymbol{S}_{7}+ \boldsymbol{S}_{6}\cdot\boldsymbol{S}_{7}
+ \boldsymbol{S}_{1}\cdot\boldsymbol{S}_{8}+ \boldsymbol{S}_{2}\cdot\boldsymbol{S}_{8}\big]\\
\quad\quad-g\mu_B B\sum\limits_{j=1}^{8} S_{j}^z+D\sum\limits_{j=1}^{8} \big(S_{j}^z\big)^2.
\end{array}
\end{equation}
The first and second parts of Eq. (\ref{Hamiltonian}) corresponds to the anisotropic Heisenberg couplings between each pair spins interacted together, which is explicitly given by
\begin{equation}
\begin{array}{lcl}
 (\boldsymbol{S}_{i}\cdot\boldsymbol{S}_{j})_{J, \Delta}=J\big(S_{i}^xS_{j}^x+S_{i}^yS_{j}^y\big)+\Delta S_{i}^zS_{j}^z,
\end{array}
\end{equation}
where $J=\{J_1, J_2\}$ denotes isotropic coupling constants, and $\Delta=\{\Delta_1, \Delta_2\}$ corresponds to the anisotropy exchange interactions. $S^\alpha$ for which ${\alpha}=\lbrace x, y, z \rbrace$ are spin-1 operators (with $\hbar=1$). In Eq. (\ref{Hamiltonian}) 
${B}$ is the applied homogeneous magnetic field in the $z$-direction. The gyromagnetic ratio would be taken as $g=2.42$ \cite{Sheikh} in the plots drawn in this paper. 
 
The characterization of the partition function of the model under consideration can be defined as
${Z}=Tr\big[\exp(-\beta H)\big],$
where $\beta=\frac{1}{k_{B}T}$, $k_{B}$ is the Boltzmann$^,$s constant and $T$ is the temperature.
The Gibbs free energy  can be obtained from the partition function of the system as $f=-k_{B}T\ln{Z}$.
 
 \begin{figure}
\begin{center}
\resizebox{0.6\textwidth}{!}{%
  \includegraphics{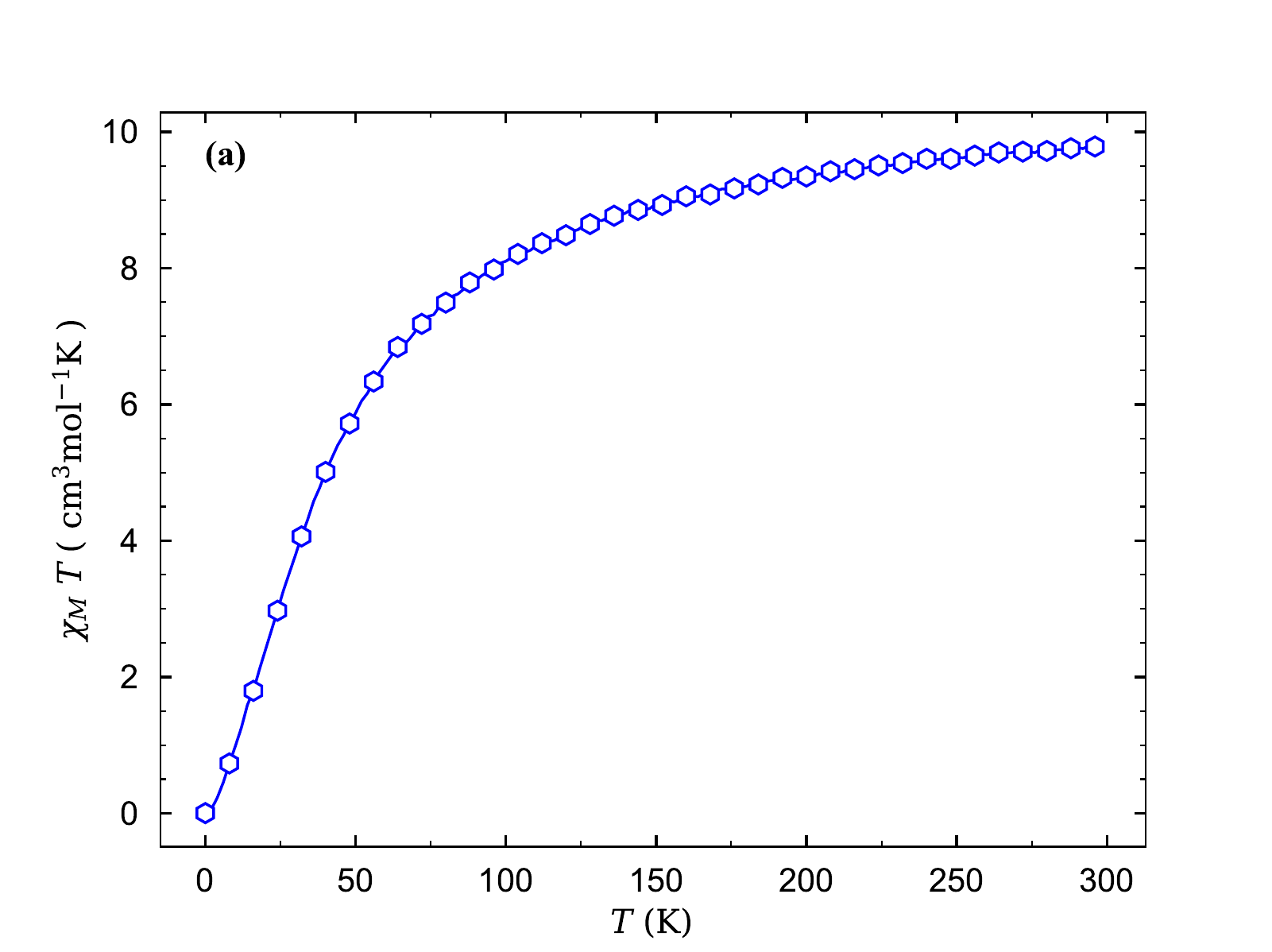}
  }
  \resizebox{0.6\textwidth}{!}{%
  \includegraphics{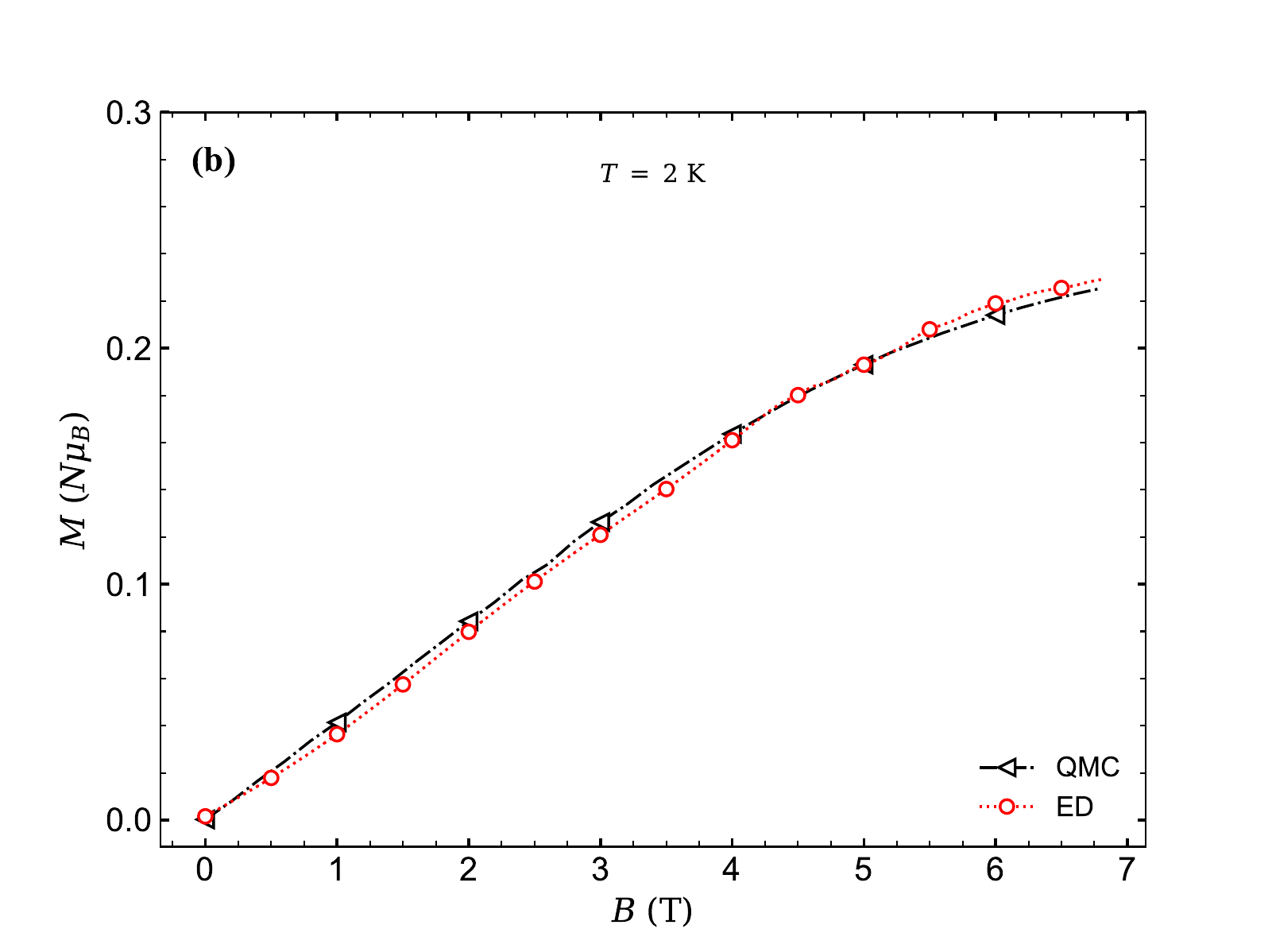}
  }
  \resizebox{0.6\textwidth}{!}{%
  \includegraphics{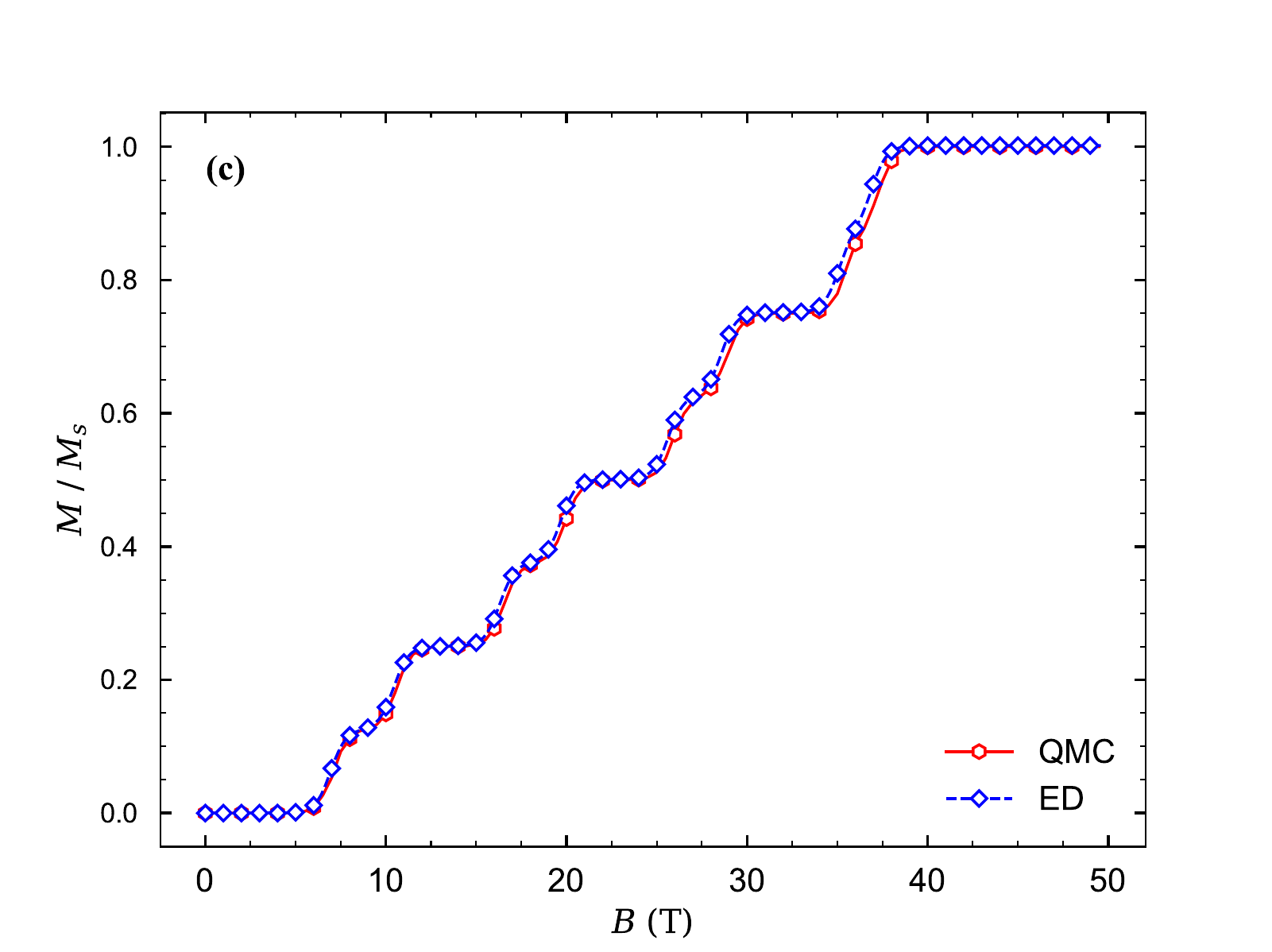}
  }
\caption{(a) QMC results for The temperature dependence of the product of magnetic susceptibility times the temperature $\chi_MT$ for the Heisenberg octanuclear nickel phosphonate-based cage in the absence of the magnetic field ($B=0$) for fixed values of $J_1=7.6\ $cm$^{-1}$ and $J_2=-22.4\ $cm$^{-1}$. Panel (b) shows deduced magnetization per saturation $M/M_s$ by QMC simulation and ED method as a function of the external magnetic field $B$ for fixed values of $J_1=7.6\ $cm$^{-1}$, $J_2=-22.4\ $cm$^{-1}$ at moderate temperature $T=2\ $K. Due to compare with the experimental data in Ref. \cite{Sheikh,Sheikh2016}, other parameters have been taken as $\Delta1=J_1$, $\Delta2=J_2$ and $D=0$. (c) The ED and QMC results for the magnetization per saturation value of the model versus magnetic field at low temperature $T=1$K for the same fixed values of other parameters as panel (a) divided by $k_{B}$, namely, $J_1/k_B=7.6\ $K and $J_2/k_B=-22.4\ $K.}
\label{fig:SuscMag_ALPS}
\end{center}
\end{figure}
 
 \section{Results and Analysis}\label{TP}
 Here we present the most interesting outcomes of our study of the magnetization process, magnetic susceptibility, and specific heat of the  butterfly-shaped octanuclear cluster $\big[\mathrm{Ni}_8(\mu_3-\mathrm{OH})_4(\mathrm{OMe})_2(\mathrm{O}_3\mathrm{PR}_1)_2 (\mathrm{O}_2\mathrm{C}^t\mathrm{Bu})_6 (\mathrm{HO}_2\mathrm{C}^t\mathrm{Bu})_8\big]$ using the standard thermodynamic relations 
\begin{equation}\label{TParameters}
\begin{array}{lcl}
{M}=-\Big(\frac{\partial f}{\partial B}\Big)_{T}, \quad {\chi}=\Big(\frac{\partial M}{\partial B}\Big)_{T}, \quad {C}=-T\Big(\frac{\partial^2 f}{\partial T^2}\Big)_{B}.
\end{array}
\end{equation}

We start with the isotropic case in Figure \ref{fig:SuscMag_ALPS}. 
Panel (a), illustrates the QMC results for the temperature dependencies of the direct current (dc) magnetic susceptibility data 
 $\chi_MT$ of the model calculated using the ALPS algorithm. 
Comparison of our numerical results with the experimental data of  Refs. \cite{Sheikh,Sheikh2016,Breeze} shows excellent agreement. 
The room temperature  value of $\chi_MT$ for the model under consideration is almost $10.0\ \mathrm{cm}^3\mathrm{mol}^{-1}\mathrm{K}$, having set parameters values to  $J_1=7.6\ $cm$^{-1}$, $J_2=-22.4\ $cm$^{-1}$ and $g=2.4$.
This is close to the expected value $9.96\ \mathrm{cm}^3\mathrm{mol}^{-1}\mathrm{K}$ for eight isolated $\mathrm{Ni}^\mathrm{II}$ ions. 
We reiterate that at this stage we are considering a pure Heisenberg spin-1 cluster system, free from any anisotropy dependencies. 
Panel \ref{fig:SuscMag_ALPS}(b) displays the ED and QMC results for the magnetization versus magnetic field at moderate temperature $T=2\ $K and the same set of coupling constants considered for examining $\chi_MT$. 
One can see from this figure that, not only ED and QMC results are in accordance with each other, but also they are in a good agreement with the experimental results reported in Ref. \cite{Sheikh,Sheikh2016}. 

Panel (c) of figure \ref{fig:SuscMag_ALPS} illustrates the  magnetization per saturation value $M/M_s$ calculated using the ED and QMC methods as a function of the magnetic field at low temperature $T=1$K for fixed values of $J_1/k_B=7.6\ $K, $J_2/k_B=-22.4\ $K and $g=2.4$. 
We mention that in order to reduce the complexity of our numerical calculations, we change the units of exchange couplings into the temperature unit (Kelvin).
It is quite clear that the magnetization curve shows intermediate plateaus at $0$, $\frac{1}{8}$, $\frac{1}{4}$, $\frac{2}{5}$, $\frac{1}{2}$, $\frac{2}{3}$, and $\frac{3}{4}$ of the saturation magnetization. 
The large number of magnetization plateaus is due to existence of a strong interdimer antiferromagnetic  interaction $J_2/k_B=-22.4\ $K. 
This figure also supports agreement between results of  both the ED and QMC approaches.

\begin{figure}
\begin{center}
\resizebox{0.45\textwidth}{!}{%
  \includegraphics{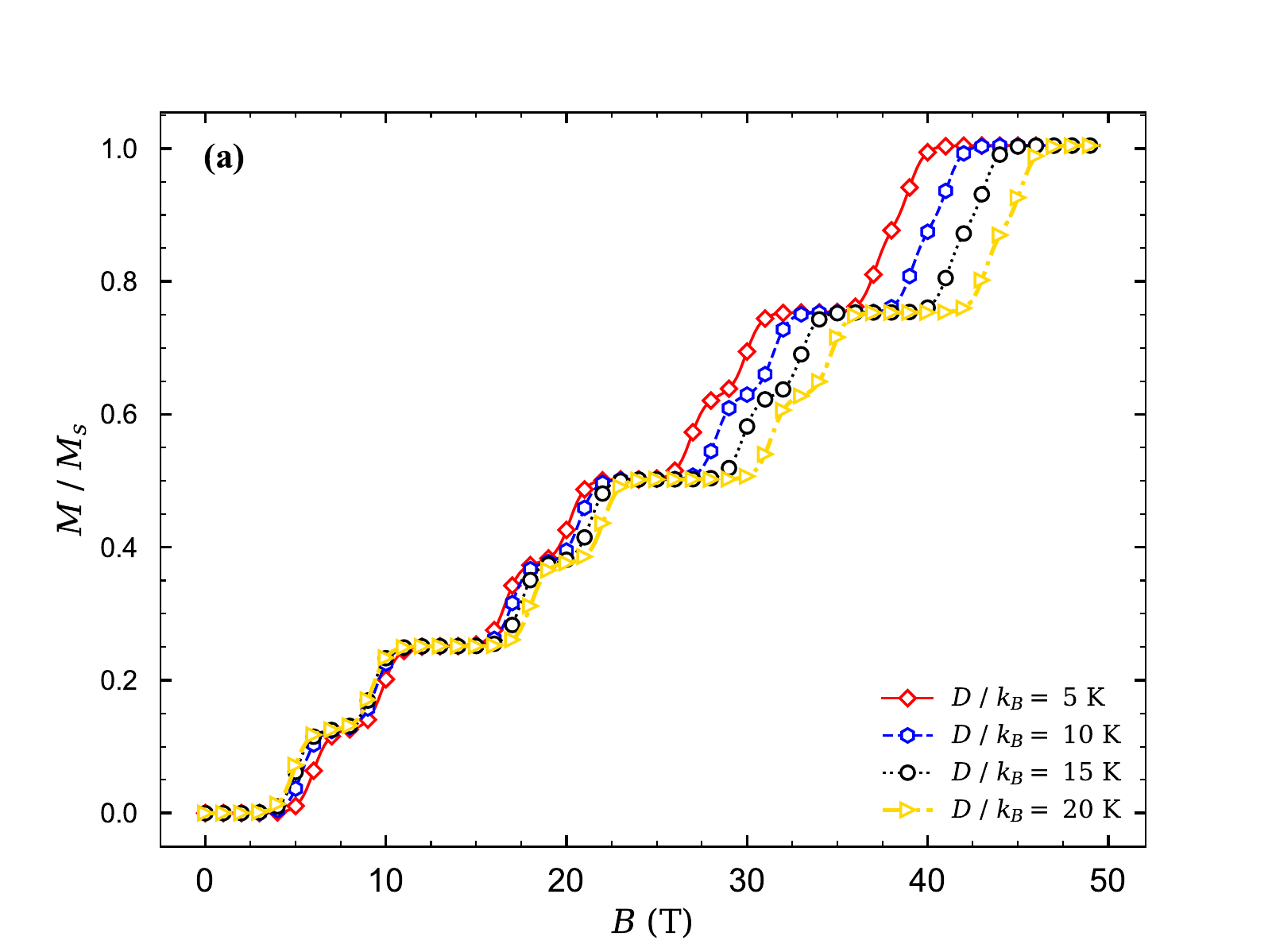}
  }
  \resizebox{0.45\textwidth}{!}{%
  \includegraphics{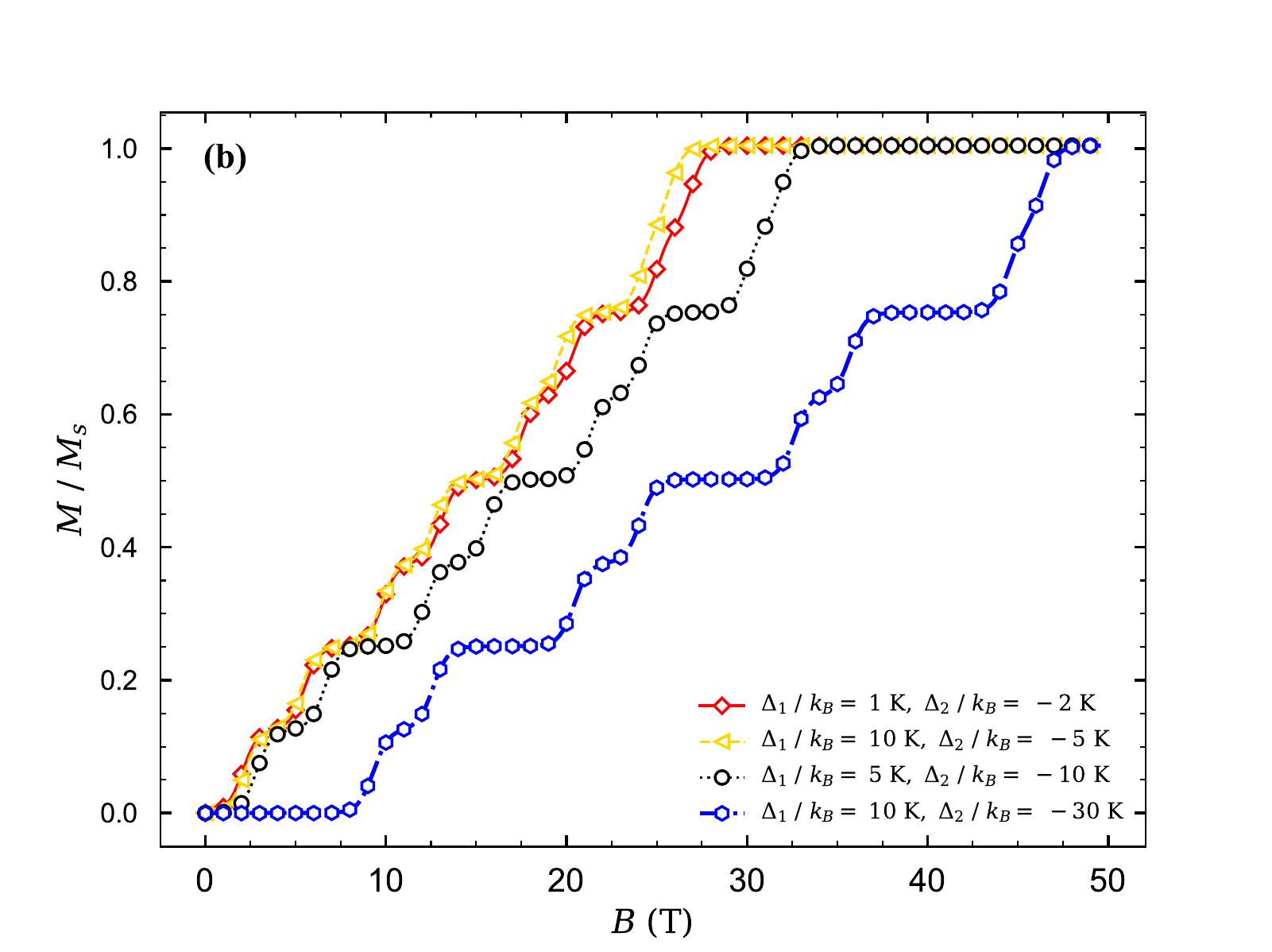}
  }
\caption{(a) ED results for the magnetization per saturation value $M/M_s$ of the isotropic octanuclear nickel phosphonate-based cage as a function of the magnetic field $B$ at low temperature $T=1$K for several fixed values of the single-ion anisotropy parameter $D$, by assuming fixed values of  $J_1/k_B=\Delta_1/k_B=7.6\ $K, and $J_2/k_B=\Delta_2/k_B=-22.4\ $K. (b) Magnetization per saturation value $M/M_s$ as a function of the magnetic field at low temperature $T=1$K for different fixed values of exchange anisotropies $\Delta_1$ and $\Delta_2$, by assuming an optional value  $D/k_B=10\ $K.}
\label{fig:Mag}
\end{center}
\end{figure}

Having established agreement between ED, QMC and experiment in the isotropic cases, we next turn our attention to the influence of anisotropy.
To this end we investigate the magnetization process and the specific heat of the model when it involves non-trivial Heisenberg exchange anisotropies $\Delta_1$ and $\Delta_2$, and the single-ion anisotropy property $D$ is turned on. 
It is considered  in Fig. \ref{fig:Mag}(a) the isotropic XXX Heisenberg case by plotting typical dependencies of the magnetization per saturation against the magnetic field $B$ for different values of the single-ion anisotropy parameter $D/k_B$,  namely $\Delta_1/k_B=J_1/k_B=7.6\ $K and $\Delta_2/k_B=J_2/k_B=-22.4\ $K. 
Interestingly, the single-ion anisotropy variations have no remarkable influence on the magnetic dependence of  magnetization in the magnetic field interval $B\lesssim 20\ $T, where a magnetization jump occurs between the $\frac{2}{5}-$plateau and  the $\frac{1}{2}-$plateau (accompanyied by the ground-state phase transition between plateaus $\frac{2}{5}$ and $\frac{1}{2}$ of the saturation magnetization). 
As a matter of fact, for the magnetic field range $B>20\ $T, intermediate magnetization plateaus at $M/M_s\geq\frac{1}{2}$ ($\frac{1}{2}$, $\frac{2}{3}$, and $\frac{3}{4}$ of the saturation magnetization) become wider and shift toward stronger magnetic fields. 
Generally, there is a delay in the magnetization behavior reaching its saturation when the single-ion anisotropy increases.

To understand the spin exchange anisotropy effects on the magnetization process of the octanuclear nickel phosphonate cage, we plot in Fig. \ref{fig:Mag}(b) the magnetization per saturation value $M/M_s$ with respect to the magnetic field for various fixed values of  both the exchange anisotropies $\Delta_1$ and $\Delta_2$, and optional value $D/k_B=10\ $K at low temperature $T=1$K. 
Coupling constants $J_1$ and $J_2$ have been taken as in panel \ref{fig:Mag}(a).
As could be expected, the width and magnetic position of all plateaus undergo significant changes upon altering anisotropies. To clarify this point,
the magnetization scenario for relatively small anisotropies $\Delta_1\ll J_1$ and $|\Delta_2|\ll |J_2|$ is illustrated in Fig. \ref{fig:Mag}(b) on a particular example with $\Delta_1/k_B=1\ $K and $\Delta_2/k_B=-2\ $K (red solid line marked with diamonds). One can see that, the width of all plateaus becomes narrower and their magnetic positions shift toward lower magnetic fields. Consequently, the magnetization reaches its saturation value in lower magnetic fields (in this case $B_s\approx 28\ $T). The similar behavior can be seen for the magnetization under condition $\Delta_1> J_1$ and $|\Delta_2|\ll |J_2|$ (orange curve marked with triangles). For the case  $\Delta_1< J_1$ and $|\Delta_2|< |J_2|$ (black dotted line with circular marks) wider magnetization plateaus are presented by magnetization curve, whose magnetic positions are in the stronger magnetic fields.

When we consider even higher degrees of exchange anisotropy  for the model such that $\Delta_1> J_1$ and $|\Delta_2|> |J_2|$, the width of all magnetization plateaus considerably increase and their magnetic positions move to stronger magnetic fields, so the magnetization reaches its saturation value in the stronger magnetic fields (for this case $B_s\approx 48\ $T).

\begin{figure}
\begin{center}
\resizebox{0.45\textwidth}{!}{%
  \includegraphics{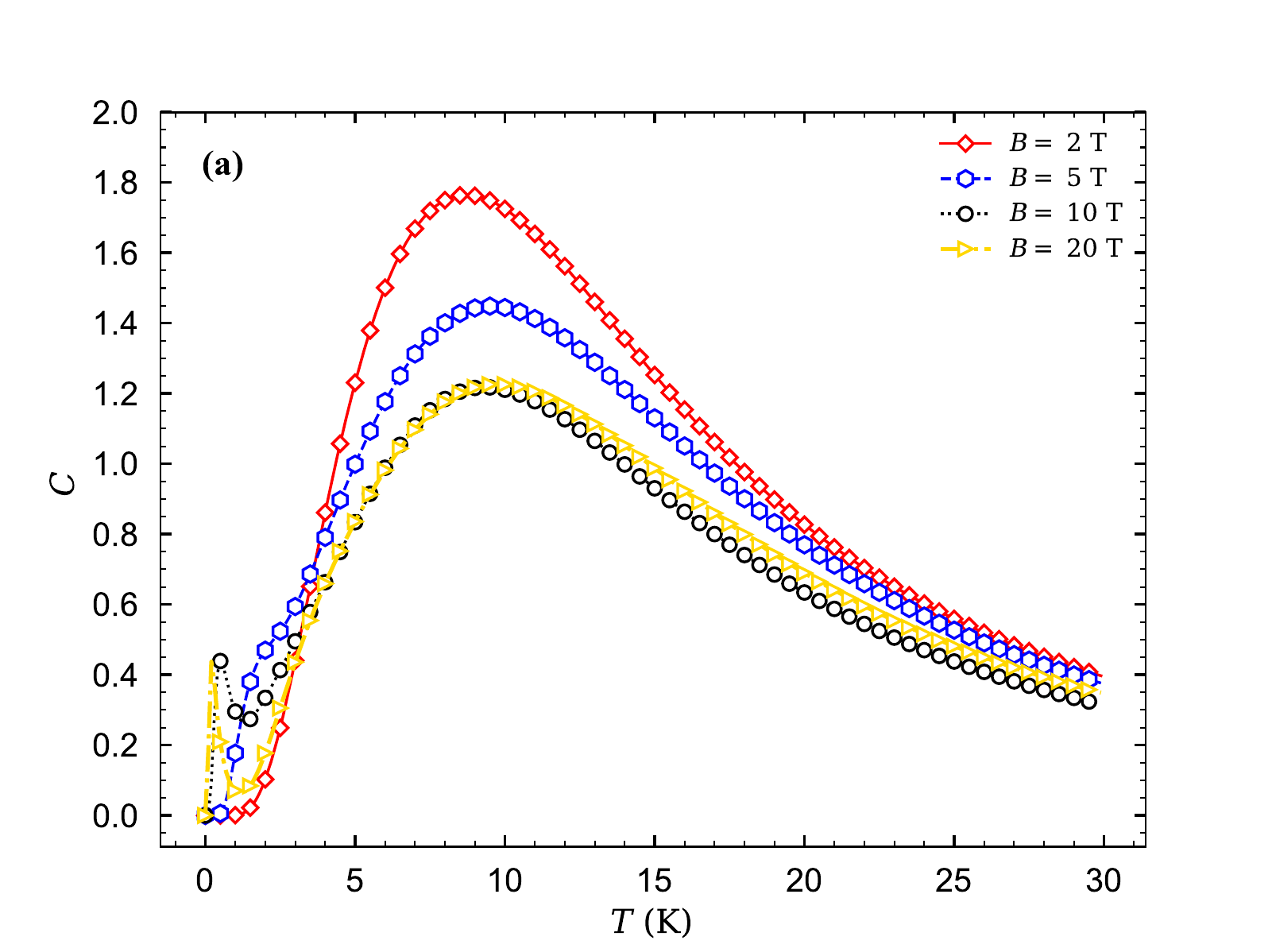}
  }
  \resizebox{0.45\textwidth}{!}{%
  \includegraphics{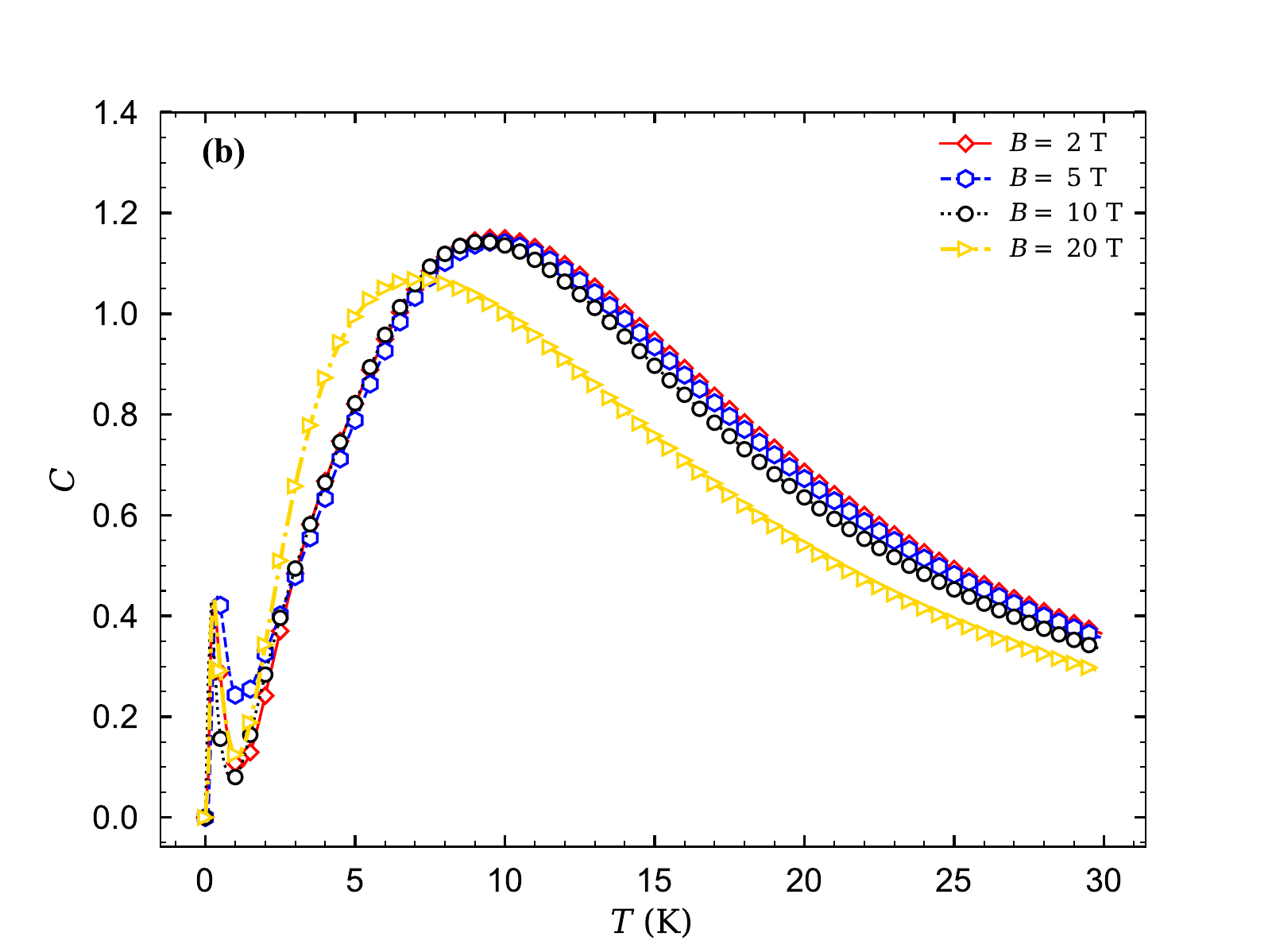}
  }
    \resizebox{0.45\textwidth}{!}{%
  \includegraphics{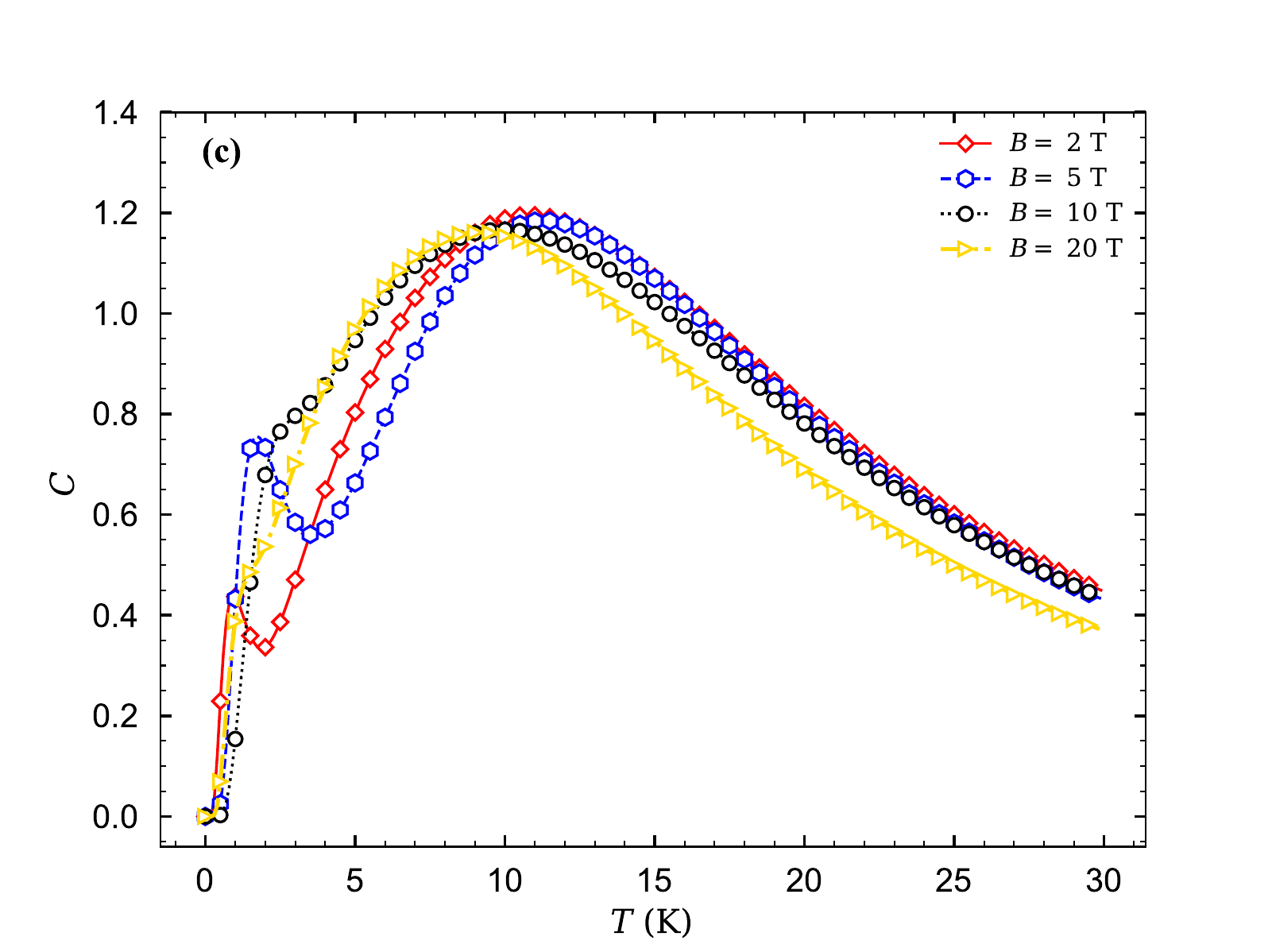}
  }
    \resizebox{0.45\textwidth}{!}{%
  \includegraphics{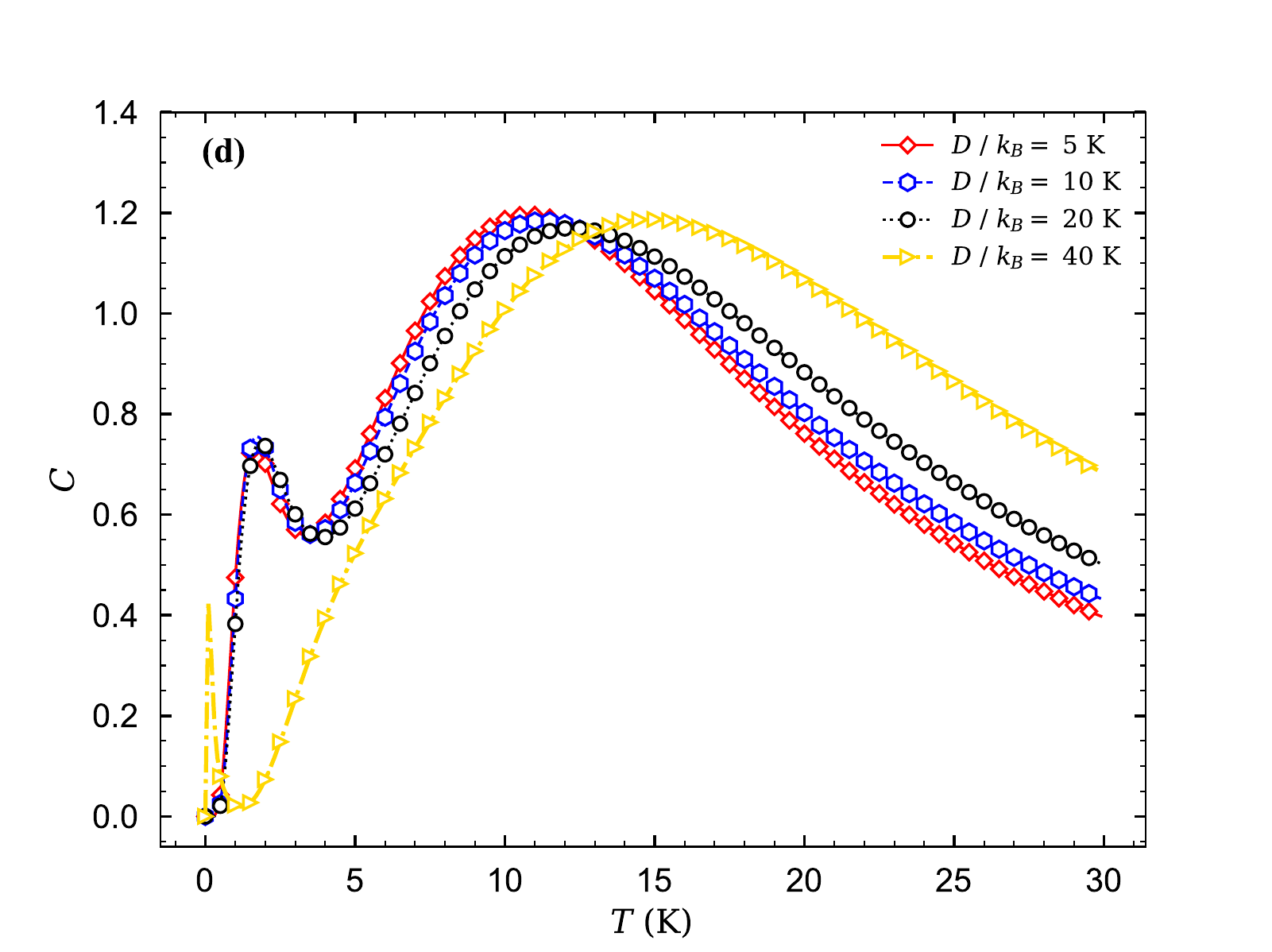}
  }
\caption{Specific heat of the isotropic Heisenberg octanuclear nickel cage as a function of the temperature for various fixed parameter values. 
Panel (a) represents the isotropic case. Here, the magnetic field is $B=2\ \mathrm{T}, 5\ \mathrm{T}, 10\ \mathrm{T}$, and $20\ \mathrm{T}$ with $J_1/k_B=7.6\ $K,  $J_2/k_B=-22.4\ $K.
The remaining parameters are $\Delta_1=J_1$, $\Delta_2=J_2$, $D=0$ and $g=2.4$. 
In panel (b) the magnetic fields are the same as in panel (a) while the anisotropies are set to 
$\Delta_1/k_B=10\ $K, $\Delta_2/k_B=-5\ $K with arbitrary single-ion anisotropy $D/k_B=10\ $K.
In panel (c) all parameters are the same as in panel (b) except the spin exchange anisotropies are set to $\Delta_1/k_B=5\ $K, $\Delta_2/k_B=-10\ $K. 
In panel (d) the magnetic field is set to the weak value of $B=5\ $T and the exchange anisotropies are the same as in panel (c) except $D/k_B=2\ \mathrm{K}, 5\ \mathrm{K}, 20\ \mathrm{K}, 40\ \mathrm{K}\ $.}
\label{fig:SHeat}
\end{center}
\end{figure}


Next, let us discuss the effects of exchange anisotropies $\Delta_1$ and $\Delta_2$ and single-ion anisotropy $D$ on the temperature dependence of the specific heat of the model for the same set of coupling constants  $J_1$ and $J_2$ as before. 
In Fig. \ref{fig:SHeat} the temperature dependence of the specific heat of the octanuclear nickel 
phosphonate-based cage under different circumstances is plotted.

In Panel~(a) this is plotted for several fixed values of the magnetic field for the isotropic Heisenberg model, namely with $\Delta_1=J_1$, $\Delta_2=J_2$, and $D=0$.
In the presence of weak magnetic fields (red and blue curves), as the temperature increases from zero, the specific heat also increases, and manifests a steep rise in the temperature interval $2\ \mathrm{K}< T<9\ \mathrm{K}\ $. It reaches its maximum at temperature $T\approx 9$K. 
Thus there is a Schottky-type maximum in the specific heat curve at $T\approx 9$K for sufficiently weak magnetic fields ($B<5\ $T). 
As the magnetic field increases further, the Schottky peak gradually decreases in height until a second peak appears at low temperature $T\approx 0.2\ $K. 
By comparing  Figs. \ref{fig:SHeat}(a) and \ref{fig:SuscMag_ALPS}(c), one sees that converting the Schottky peak into a double-peak coincides the magnetization jump from zero plateau to first intermediate $\frac{1}{8}-$plateau.

The effects of anisotropy are demonstrated in the remaining panels of Fig. \ref{fig:SHeat}.
In panel~(b) the temperature dependence of the specific heat for the anisotropic case where $\Delta_1/k_B=10\ $K, $\Delta_2/k_B=-5\ $K and $D/k_B=10\ $K is plotted for  several fixed values of the magnetic field $B$. 
We observe a stable double-peak temperature dependence for each of these values. 
For  weak magnetic fields ($B=2\ \mathrm{T}, 5\  \mathrm{T}, 10\  \mathrm{T}$) the specific heat curves are very similar. 
But when the the magnetic field increases beyond $B=10\ $T, the height and the position of larger peak changes; its height decreases as does its peak temperature. 

Fig. \ref{fig:SHeat}(c) illustrates that, when still different anisotropies are considered for the model, the double-peak gradually changes back to the Schottky peak at lower temperatures upon increasing the magnetic field. 
This scenario indicates that the system returns to the ground-state phase corresponding to the magnetization $\frac{1}{2}-$plateau.

Finally, our exact
results for the effects of single-ion anisotropy on the temperature dependence of the specific heat are plotted in Fig. \ref{fig:SHeat}(d), where other parameters have been taken as $\Delta_1/k_B=5\ $K, $\Delta_2/k_B=-10\ $K and $B=5\ $T.
Interestingly, increase of the single-ion anisotropy results in altering the shape and the temperature position of the larger peak of the double-peak appeared in the specific heat curve. Strong single-ion anisotropy property has strong influence on the height and temperature position of the both peaks. 
Generally, both peaks move away from each other on the temperature axis. 
Indeed, by increasing the single-ion anisotropy, the specific heat appears like a single Schottky maximum at higher temperatures, while there is still a cusp reminiscent of the smaller peak at sufficiently low temperatures.
These changes in specific heat behavior coincide with the magnetization response to the single-ion anisotropy variations (see Fig. \ref{fig:Mag}(a)).

\section{Conclusions}\label{conclusions}
In this paper, we have theoretically investigated the magnetic and thermodynamic properties of the octanuclear nickel phosphonate-based cage with the geometry of butterfly-shaped molcular structure. To this end, we first examined the magnetization processes of the isotropic version of the  model through ED and QMC methods. 
Our results are in  excellent agreement with the experimental data. 
Building on this confidence, we continued our investigations by means of the ED procedure for the magnetization process, as well as, the specific heat of the model including Heisenberg exchange anisotropy and single-ion anisotropy properties.

According to our investigations, the model has a complex magnetization landscape with a number of intermediate plateaus and magnetization jumps at low temperature accompanying with the ground-state phase transitions. 
As a result, we found that the magnetization behavior strongly depends on the Heisenberg exchange anisotropy and single-ion anisotropy properties. 
There is a clear competition between spin exchange anisotropies and the single-ion anisotropy to alter the width and magnetic position of the magnetization plateaus. 
Strong single-ion anisotropy leads to an increase in  the width of intermediate magnetization plateaus, also a shift in their magnetic positions toward stronger magnetic fields.

Furthermore, it has been demonstrated that for the isotropic case, the specific heat manifests a Schottky-type maximum in the temperature position $T\approx 9\ $K when the system is placed in the presence of  weak magnetic fields. 
The shape and temperature position of the Schottky peak are strongly dependent on the magnetic field and strong fields induce a second peak at finite low temperatures. 
By imposing anisotropies in the Hamiltonian of the model, we uncovered that, in addition to the magnetic field, anisotropy parameters play an important role for  the height and  position of all peaks appeared in the specific heat curve. 
Moreover, by comparing the low-temperature magnetization behavior and the temperature dependence of the specific heat, we have understood that the specific heat changes made by the anisotropies alterations are in accordance with the magnetization jumps, indicating the ground-state phase transition.

We consider these results interesting enough to motivate further studies on  the effects of anisotropies  on the cooling/heating and magneticaloric process of the octanuclear nickel phosphonate-based cage and/or similar small spin clusters. 
This will be the subject of our future works.

\section*{Acknowledgments}
H. Arian Zad and N. Ananikian acknowledge the receipt of the grants from the ICTP Affiliated Center Program AF-04 and the CS MES RA in the frame of the research project No. SCS 18T-1C155.   
 The authors are also grateful to prof. M. Ja\v {s}\v{c}ur , J. Stre\v{c}ka, H. Poghosyan, and 
V. Abgaryan  for their useful discussions.

\bibliography{}

\end{document}